\documentstyle[twoside,fleqn,espcrc2,epsfig,german]{article}

\newcommand{\AmS}{{\protect\the\textfont2
  A\kern-.1667em\lower.5ex\hbox{M}\kern-.125emS}}

\title{Finite Temperature Phase Diagram of QCD with improved Wilson
  Fermions \thanks{Work in part supported by NATO grant no. CRG940451 and by
  the  European Community TMR Programs TRACS and ERBFMRX-CT97-0122}}

\author{ M. Oevers \address{Fakult"at f"ur Physik, Universit"at Bielefeld,
    Postfach 100131, 33501 Bielefeld, Germany} \address{Dept. of Physics and 
Astronomy,University of Glasgow G12 8QQ, U.K.} with 
F. Karsch$^{{\footnotesize{a}}}$ , 
E. Laermann$^{{\footnotesize{a}}}$, 
P. Schmidt$^{{\footnotesize{a}}}$ }

\begin{document}
\pagestyle{empty}
\setcounter{topnumber}{1}
\begin{abstract}
We present results of an ongoing study of two flavour QCD with Wilson fermions
at finite temperature. We have used tree level Symanzik improvement in both 
the gauge and fermion part of the action. The phase diagram was previously 
determined on an $8^3 \times 4$ lattice and the existence of Aoki's phase 
demonstrated. On our current lattice of $12^2 \times 24 \times 4$ we have 
extended the set of observables and studied chiral and thermal properties at 
light quark masses.
\end{abstract}
\maketitle
\section{Introduction}
The study of the finite temperature phase diagram of 2 flavour QCD with Wilson
quarks has revealed a rather intricate picture. 
As was pointed out already some time ago \cite{Aoki-I}, the pion can become
massless despite the fact that the Wilson term breaks chiral symmetry, because
of the existence of a second order phase transition into a parity and flavour
symmetry violation phase. At finite temperature this
phase is expected to pinch into a cusp that meets with the $m=0$ line coming in 
from weak coupling \cite{Phdiag}. The thermal line of the deconfinement phase 
transition crosses the $m_q=0$ line and in \cite{Creutz} it was argued that it
should bounce back towards weaker coupling due to a symmetry under the change of
sign of the mass term in the continuum theory . The relation of this crossing
point to the tip of the cusp of the Aoki phase is somewhat subtle. First of all
the thermal line should not run into the Aoki phase, since this would mean, that
one has a massless pion in the deconfined phase. The thermal line can now
either touch the cusp of the Aoki phase or not. In the latter case one would
have a transition line (probably first order) from a confined phase with $m_q >
0$ to a confined phase with $m_q < 0$, without the pion actually becoming
massless  \cite{ShaSin}. 
\section{Results}
Utilizing a tree level $O(a^2)$ improved action for the gauge fields and the
tree level clover action for the fermion fields, we have, in a previous study on
an $8^3\times4$ lattice, mapped out the phase diagram using the pion norm and the
Polyakov loop as observables. In the present study we have increased the lattice
size to  $12^2 \times 24 \time 4$ to check finite size effects and to measure the
pion screening mass and quark mass accurately. From these we can also determine
the properly subtracted order parameter of chiral symmetry breaking for Wilson
fermions.
\subsection{Quark mass}
The current quark mass is defined via the axial Ward identity \cite{ChWI}:
\begin{equation}
      2m_q \equiv \frac{\nabla_\mu \langle 0 |  A^\mu | \pi \rangle}
      {\langle 0 | P | \pi \rangle}
\end{equation}
Our results are shown in figure 1. For $\beta=2.8$ we find some curvature for
the quark mass as a function of $1/\kappa$, though a linear fit produces a
reasonable $\chi^2$. For $\beta=3.1$ we have also explored a region of hopping
parameters where the quark mass becomes negative. There we find a rather
peculiar behaviour. The quark mass decreases as one lowers $\kappa$ towards 
$\kappa_c$ (see section \ref{kappa_c}), which is in contrast to simulations with
unimproved Wilson fermions that have shown the same qualitative behaviour for 
positive and negative quark masses \cite{Phdiag}.
\begin{figure}[tl]
  \epsfig{file=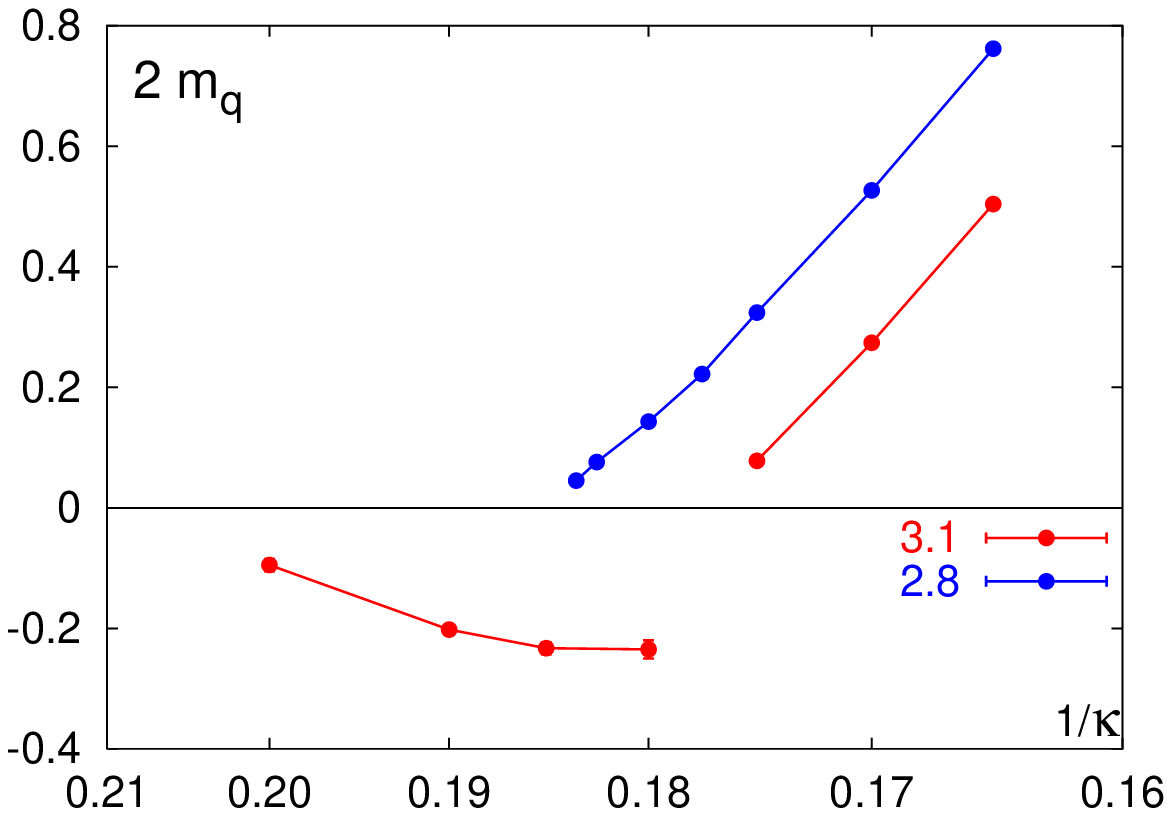, width=7.5cm}
  \begin{minipage}[t]{7.5cm}
    {Figure 1. Current quark mass $2 m_q^{WI}$ as a function of
      $1/\kappa$ for $\beta = 2.8$ and $\beta = 3.1$. }   
  \end{minipage}
\end{figure}
\subsection{Pion screening mass}
We have also accurately measured the pion screening mass. Our results are
depicted in figure 2. For $\beta=2.8$ the decrease of the pion mass is consistent
with a linear behaviour $m_\pi^2 \propto 1/\kappa - 1/\kappa_c$ down to small
values of $m_\pi^2$. This also applies for $\beta=3.1$ for $\kappa$ values that
correspond to positive quark masses. For negative quark masses the behaviour is
quite different and inconsistent with a linear behaviour for the points 
measured. We will argue below that for $\beta=3.1$ the pion mass does not go to
zero as the quark mass goes to zero, because one crosses the finite temperature
transition line before the quark mass becomes zero. 
\begin{figure}[tr]
  \epsfig{file=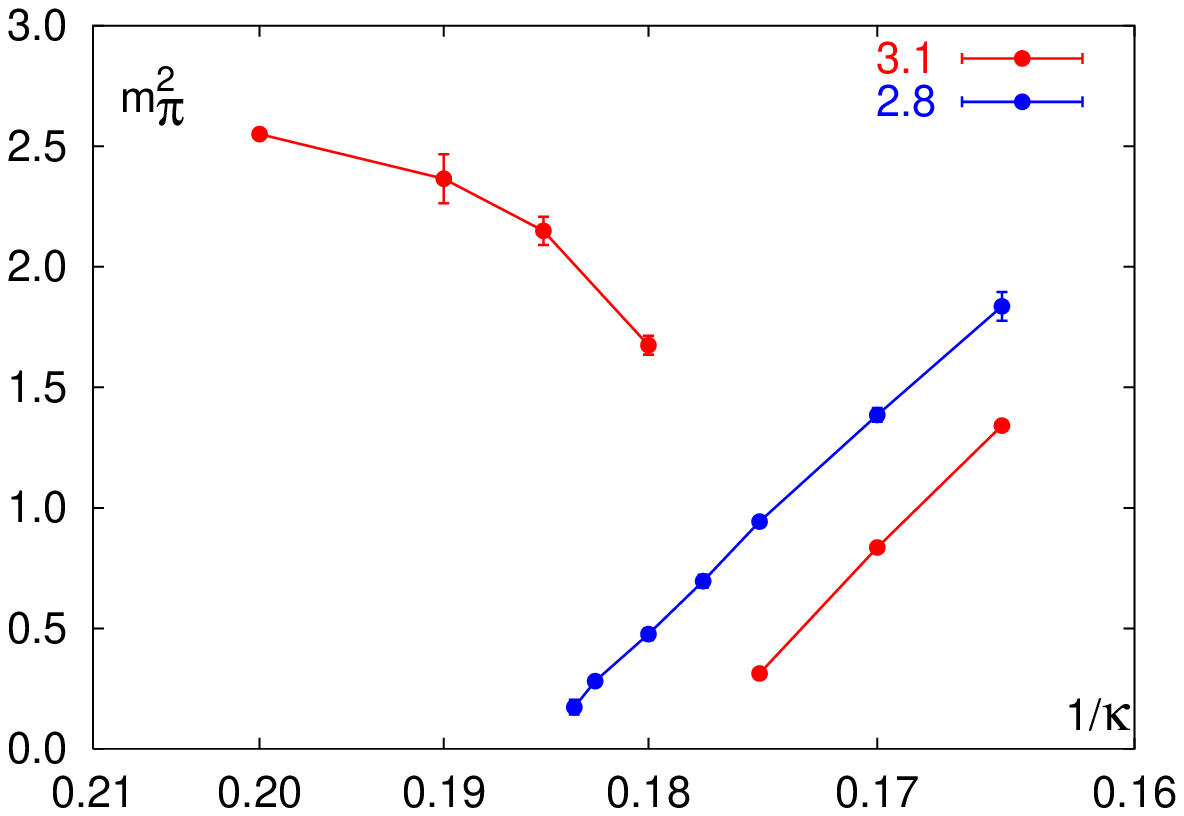, width=7.5cm}
  \begin{minipage}[t]{7.5cm}
    {Figure 2. Pion screening mass squared as a function of $1/\kappa$
      for $\beta = 2.8$ and $\beta = 3.1$.}  
  \end{minipage}
\end{figure}
\subsection{Locating the critical line}
\label{kappa_c}
To locate the critical hopping parameter $\kappa_c$ we have performed the
following fits to our data. We have fitted $m_\pi^2$ linearly and $m_q$ 
quadratically as a function of $1/\kappa$. Since the pion norm near $\kappa_c$ 
is expected to behave like $\Pi \approx 1/m_{\pi}^2$, we have also extracted
a value for $\kappa_c$ from this observable. Since the data show quite some
nonlinear behaviour as a function of $1/\kappa$, we have used a quadratic
fit ansatz.  
As the results from the different fits turned out to be consistent, we took
confidence in using our data for the pion norm from our simulation at
$\beta=3.0$ on the smaller lattice to get an estimate of $\kappa_c$ for
$\beta=3.0$. Our results are summarized in the following table:
\begin{center}
\begin{tabular}{|c||c|c|c|} \hline
$\kappa_c$ & $ \beta= $2.8 & $ \beta= $3.0 & $ \beta= $3.1 \\ \hline
$m_\pi$    & 0.1860(2)     &               & 0.1783(2)     \\ \hline
$m_q$      & 0.1853(3)     &               & 0.1770(3)     \\ \hline
$\Pi$      & 0.1859(3)     & 0.1823(10)    & 0.1800(5)     \\ \hline
\end{tabular}
\end{center}
\subsection{Chiral order parameter}
Because of the explicit breaking of chiral symmetry by the Wilson action, one
has to define a properly subtracted order parameter to obtain the correct
continuum limit. Using axial Ward identities the order parameter is defined as
follows\cite{ChWI}:
\begin{equation}
    \langle\bar{\psi} \psi\rangle_{sub} = 2m_q \cdot Z \cdot 
    \sum_{x}\langle\pi(x)\pi(0)\rangle
\end{equation}
Here $Z$ is a renormalization factor for which we take its tree level value 
$Z=(2\kappa)^2$. The sum over the pion correlation function is just the pion
norm. For $\beta=2.8$ the data extrapolate to a finite intercept at $m_q=0$. 
For $\beta=3.1$ the data show more curvature and we expect 
$\langle\bar{\psi} \psi\rangle_{sub}$ to go to zero, though a definitive
statement has to await further data at smaller quark masses.
\begin{figure}[tl]
  \epsfig{file=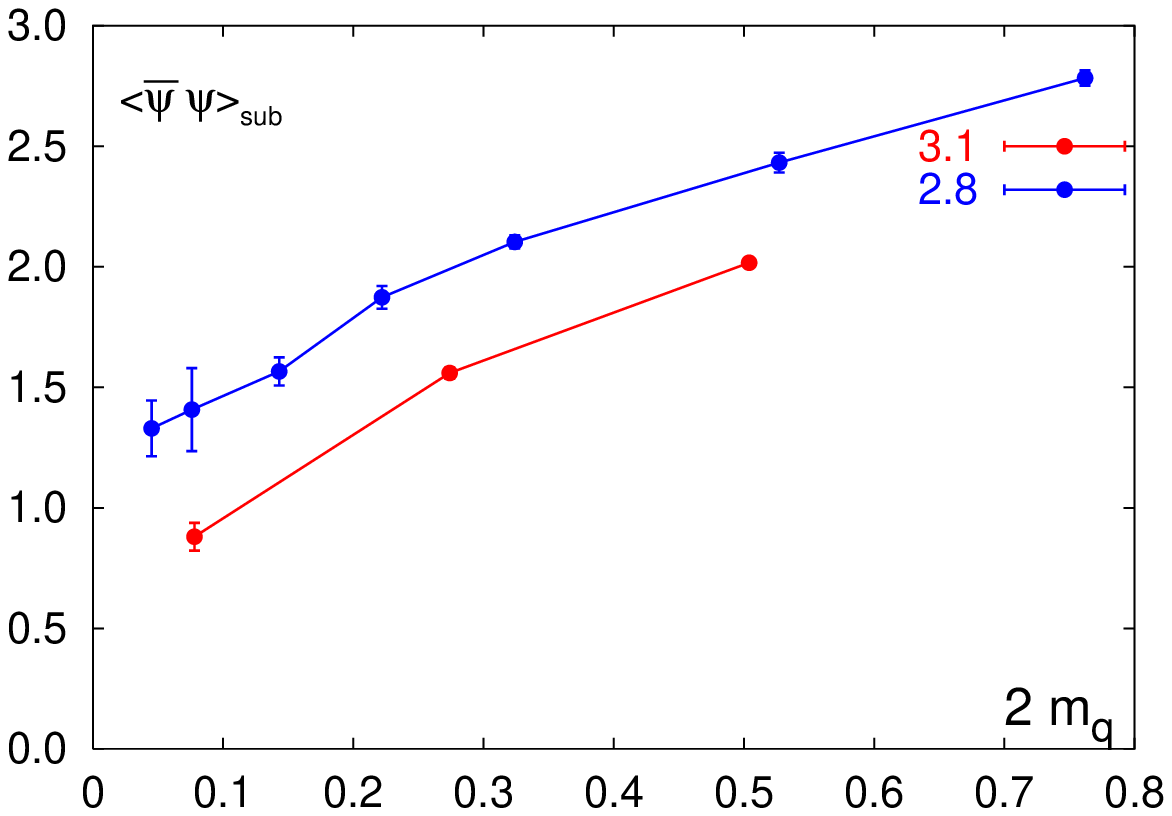, width=7.5cm}
  \begin{minipage}[t]{7.5cm}
    {Figure 3. Subtracted chiral condensate as a function of $2 m_q$ for
      $\beta = 2.8$ and $\beta = 3.1$.}
  \end{minipage}
\end{figure}
\subsection{Polyakov loop}
Figure 4 shows our results for the Polyakov loop as a function of $\kappa$
including data from both lattices. The vertical lines indicate the position
of the extrapolated $\kappa_c$, which decreases as $\beta$ increases. 
For $\beta=2.8$ all our data for the pion mass lie in the confined region. We
therefore conclude that the pion mass vanishes as the quark mass goes to
zero, i.e. for $\beta=2.8$ we hit the Aoki phase as we increase $\kappa$. For
$\beta=3.0$ and $3.1$ this is no longer so clear. The Polyakov loop already 
shows a high temperature behaviour where the extrapolated pion mass would be 
small. For $\beta=3.0$ the situation is less prominent and
one could still argue that the pion becomes massless, but since the Polyakov
loop is in the high temperature phase immediately after one crosses $\kappa_c$, 
the finite temperature line and the line of vanishing quark mass come very close
together. For $\beta=3.1$ it becomes evident, that one crosses the finite
temperature transition line before the line of vanishing quark mass. The pion
will therefore not become massless as the quark mass vanishes. 
Because the Polyakov loop continues to rise past the $m_q=0$ line, one can
exclude that the finite temperature line bounces back towards weaker
coupling immediately. 
\begin{figure}[tr]
  \epsfig{file=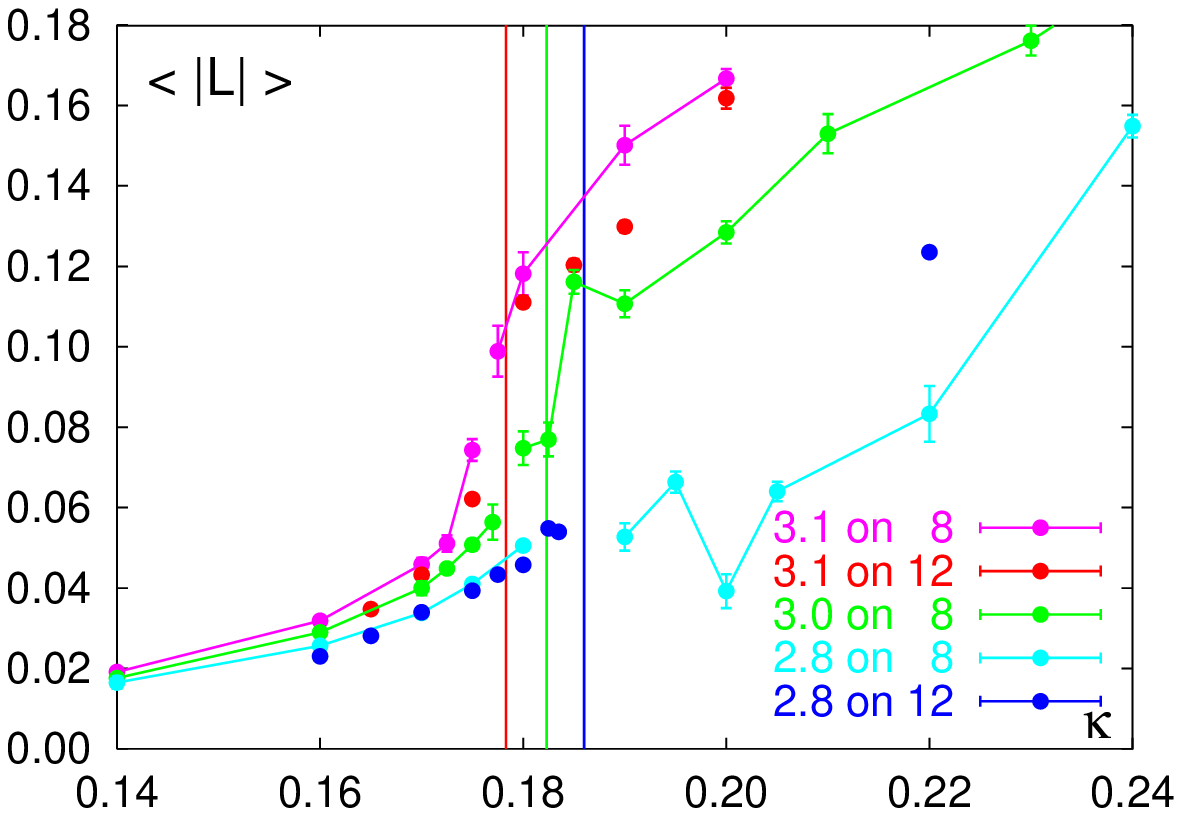, width=7.5cm}
  \begin{minipage}[t]{7.5cm}
    {Figure 4.  Polyakov loop as a function of $\kappa$ for $\beta = 2.8$,
      $\beta = 3.0$ and $\beta = 3.1$ }  
  \end{minipage}
\end{figure} 
\section{Summary and Conclusions}
From the vanishing of the pion mass at zero quark mass we conclude
that at $\beta=2.8$ there exists a parity and flavour symmetry breaking phase. 
At $\beta=3.0$ the finite temperature transition line and the
line of vanishing quark mass come at least very close and for $\beta=3.1$ the
Aoki phase ceases to exist and one crosses the finite temperature transition
line before the line of vanishing quark mass and therefore the pion does not
become massless. To make these claims even stronger it would be desirable to
lower the quark mass further for $\beta=3.1$ and to see the pion mass rising
again. Once the phase diagram is known one can start to investigate the
thermodynamic behaviour at small pion masses.


\begin{thebibliography}{9}
\bibitem{Aoki-I}  S. Aoki, Phys. Rev. D30 (1984) 2653;
                           Phys. Rev. Lett. 57 (1986) 3136;
                           Nucl. Phys. B314 (1989) 79.
\bibitem{Phdiag}  S. Aoki, A. Ukawa, T. Umemura, 
                           Phys. Rev. Lett. 76 (1996) 873;
                           Nucl. Phys. B (Proc. Suppl.) 47 (1996) 511;
                  S. Aoki {\it et. al.}, 
                           Nucl. Phys. B (Proc. Suppl.) 53 (1997) 438;
                  S. Aoki, T. Kaneda, A. Ukawa, 
                           Phys. Rev. D56 (1997) 1808.
\bibitem{Creutz} M. Creutz, hep-lat/9608024
\bibitem{ShaSin} S. Sharpe, R. Singleton Jr., hep-lat/9804028
\bibitem{ChWI}   M. Bochicchio {\it et. al.}
                           Nucl. Phys. B262 (1985) 331;
                 S. Itoh {\it et. al.}
                           Nucl. Phys. B274 (1986) 33.

\end{thebibliography}
\end{document}